# WhatsAI: Transforming Meta Ray-Bans into an Extensible Generative AI Platform for Accessibility

Nasif Zaman[1], Venkatesh Potluri[2], Brandon Biggs[1] and James M. Coughlan[1]

[1]Smith-Kettlewell Eye Research Institute, [2]University Of Michigan

## Abstract

Multi-modal generative AI models integrated into wearable devices have shown significant promise in enhancing the accessibility of visual information for blind or visually impaired (BVI) individuals, as evidenced by the rapid uptake of Meta Ray-Bans among BVI users. However, the proprietary nature of these platforms hinders disability-led innovation of visual accessibility technologies. For instance, OpenAI showcased the potential of live, multi-modal AI as an accessibility resource in 2024, yet none of the presented applications have reached BVI users, despite the technology being available since then. To promote the democratization of visual access technology development, we introduce WhatsAI, a prototype extensible framework that empowers BVI enthusiasts to leverage Meta Ray-Bans to create personalized wearable visual accessibility technologies. Our system is the first to offer a fully hackable template that integrates with WhatsApp, facilitating robust Accessible Artificial Intelligence Implementations (AAII) that enable blind users to conduct essential visual assistance tasks, such as real-time scene description, object detection, and Optical Character Recognition (OCR), utilizing standard machine learning techniques and cutting-edge visual language models. The extensible nature of our framework aspires to cultivate a community-driven approach, led by BVI hackers and innovators to tackle the complex challenges associated with visual accessibility.

## 1. Introduction: The Untapped Potential of Hands-Free Vision Assistance

Camera-equipped wearable technology, exemplified by Meta Ray-Bans (MRB), enables promising new possibilities for blind and low vision users. Unlike traditional smartphone applications requiring manual camera aiming, which can be challenging for users who can't clearly see the camera viewfinder, MRB offers a more natural experience where users can direct the camera simply by orienting their head, using proprioceptive cues and responding to voice commands to frame scenes of interest. This hands-free capability significantly lowers barriers to accessing visual information, making tasks such as reading signs, identifying objects, and navigating unfamiliar environments more seamless and less conspicuous (Waisberg et al. 2024). A blind co-author recounts an actual sample usage of MRB: "One very transformational experience I've been having is during airport air travel. I previously relied on assistance and just during my flight to CSUN, I completely ignored the airport assistance and went through security to my gate using Aira on MRB in under 40 minutes with a pickup of a coffee on the way, and that was wildly empowering. That's amazing."

Recent research by Seiple et al. (2025) highlights the importance of AI-powered assistance for people with vision impairment. Their study comparing various Assistive Artificial Intelligence Implementations (AAIIs) found that AI tools significantly improved users' ability to complete various daily tasks compared to baseline conditions without AI assistance, with particularly strong performance improvements for text-based tasks. Similarly, Adnin and Das (2024) found that blind individuals incorporate GenAI tools into a wide variety of content creation and information retrieval tasks, ranging from preparing written materials to programming and visual question answering. Further, Herskovitz et al. (2023) show that BVI users wish to create custom visual accessibility technologies, and presented a no-code interface to equip BVI users to build their own visual access workflows. Among blind users, MRB has grown in popularity (Stern 2025). This has prompted Meta to integrate Be My Eyes to provide better service to the blind community (Meta 2024) and Aira, a popular visual interpreting service to provide service via the Ray-Bans through WhatsApp (Aira 2025). An appealing aspect of this wearable is the inconspicuous nature of the smart hardware and the fashionable design of the frame.

However, the current MRB ecosystem operates as a largely closed platform, lacking open APIs that would enable developers and users to extend its functionalities for specific accessibility needs. While the integrated Meta AI offers baseline capabilities, its closed nature limits innovation. WhatsAI confronts this limitation by offering an open and extensible template that harnesses the power of generative AI models.

## 2. An Extensible AAII Framework

Our project provides a comprehensive framework for building AAII systems accessed remotely via WhatsApp. The architecture (see Fig. 1) comprises the following elements:

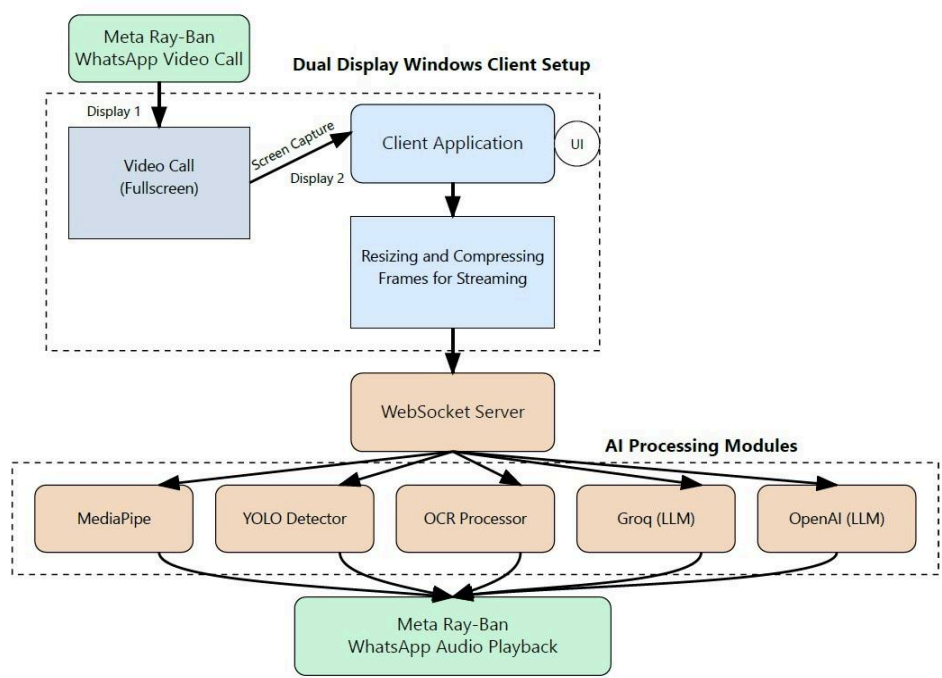

Figure 1. AAII: Accessible Artificial Intelligence Implementation Architecture.

### 2.1 WhatsApp Integration

WhatsApp is a popular chat/video calling app. The WhatsApp integration facilitates remote access to AI-powered vision assistance. A user wearing MRB or using a smartphone initiates a WhatsApp video call to a computer running our client application. The user switches the call's camera feed from the phone to the MRB by double-pressing the side button. The client captures the video feed displaying the focused participant and streams it to a local or remote server for processing.

## 2.2 Client-Server Architecture

The client-server architecture ensures flexible deployment and resource utilization. In this pipeline, the client software does not need to be in the same operating system as the server. Therefore, multiple AI services can be running in the server while the client workflow can be customized without breaking the service for others. Moreover, servers can be hosted on cloud GPU while clients can be run on any Windows computer. The client, implemented using Python with libraries like tkinter for the GUI and MSS for screen capture, handles video frame transmission. It also transmits a processor ID that specifies how the server should process the frame. The server, built using FastAPI and Uvicorn, processes these frames using ProcessorManager to switch between various AI models (Fig. 1) and returns processed information, including visual overlays and textual descriptions. The client then plays the text description using Text to Speech which is transmitted to the Meta Ray-Bans using virtual audio cables.

## 2.3 Extensible Processing Options

Our template is explicitly designed for extensibility and customization, empowering both users and developers to adapt the system to their unique needs. The BaseProcessor class serves as the foundation for implementing new video stream processing capabilities. Developers can create custom processors by inheriting from this class and overriding the process_frame method to integrate different AI models or algorithms. Some processors are implemented as examples:

- **Dense Region Captioning and OCR:** Leveraging the Florence-2 model for detailed image captioning and text recognition
- **YOLO Detection:** Utilizing YOLOv11 for real-time object detection and segmentation
- **MediaPipe:** Enabling face, hand, and body pose estimation
- **Groq:** Providing live image description using llama-3.2-90b-vision-preview for low-latency inference
- **OpenAI:** Offering image understanding through GPT-4o

This modular design enables various customizations. For instance, a user could create a processor that leverages an OCR engine to find a specific packaged food in a grocery store. To that end, they would need to extend the BaseProcessor and create FindItemProcessor in the server/processor folder and override the process_frame so that if the recognized text matches with the search term, an egocentric direction is returned. This class will need to be imported in stream.py and listed in the cls.processors dictionary inside the ProcessorManager. This processor_id will need to be added to the client processor dropdown menu, where the user can select it from among multiple options.

## 2.4 GitHub Repository

The entire project, including comprehensive setup instructions, is available as an open-source repository on GitHub (https://github.com/Znasif/HackTemplate), encouraging community contributions and customization.

# 3. Usefulness for Blind Users: Addressing Key Accessibility Needs

Research by both Seiple et al. (2025) and Adnin and Das (2024) demonstrates the significant benefits AI-powered vision assistance can provide to blind users. For example, Adnin and Das

(2024) discovered that blind individuals found significant value in GenAI tools for various information-seeking tasks. Visual question answering tools like Be My AI provided more detailed and systematic descriptions than those from sighted people, offering rich information about foregrounds, backgrounds, attire, objects, colors, and overall ambiance. Participants appreciated GenAI for increasing their workflow efficiency by eliminating hours of searching, combining, revising, reformatting, and proofreading information. Additionally, these tools helped them develop new skills and enhance proficiency in areas where they previously struggled, allowing them to receive feedback and explanations without feeling judged for asking "stupid questions." The applications ranged from songwriting and understanding academic jargon to exploring mathematical concepts and practicing programming, demonstrating how GenAI served as both an accessibility tool and a learning companion.

Seiple et al. (2025) revealed significant differences in the effectiveness of various AAII for people with vision loss. For text-based tasks, Envision, Seeing AI, and Lookout substantially improved task completion rates (with significantly higher odds ratios for 100% of text tasks), while OrCam showed improvement in 83% of tasks. For non-text searching and identification tasks, Seeing AI demonstrated the highest effectiveness (83% improved odds ratios), followed by both OrCam and Envision (50%), and finally Lookout (33%).

These results demonstrate that blind users are already effectively using AAII in their daily life. However, most of these implementations work on static images and don't allow instant live feedback. Wearable options like Android XR Gemini live and Meta AI live are still not publicly available and don't have any customizability. Our template leverages these insights to address key accessibility challenges by empowering blind users to customize the tools for themselves:

## 3.1 Hands-Free Operation, Low-Latency AI Assistance

The integration with Meta Ray-Bans provides a hands-free means of capturing visual information, a feature blind users value for its natural and discreet interaction (Adnin & Das, 2024). Additionally, the use of Groq's llama-3.2-90b-vision-preview ensures low-latency live descriptions, enabling users to quickly interpret dynamic environments and engage more fluidly with the world. Individuals can extend the prompts for their specific use case. While exploring a new side of town, the prompt could be "read out loud any new building name you recognize" - which may help them know what kinds of food options and attractions are available nearby while keeping their hands free.

## 3.2 Versatile AI Capabilities for Daily Tasks

Our platform offers multiple AI-driven functions, including live description, object detection (e.g., to help the user find their keys on a crowded table (Morrison et al. 2023)), and OCR, allowing users to access visual information based on their needs. Seiple et al. (2025) found that such AI tools significantly improve blind users' ability to complete daily tasks, especially text-related ones. By leveraging these capabilities, we enhance independence and ease of access to important information.

Our platform offers multiple AI-driven functions that assist blind individuals in their daily lives with specific applications. The live description feature narrates crowded environments like coffee shops, detailing table arrangements and available seating options. Object detection helps users locate misplaced items such as keys on a cluttered desk, AirPods on a bookshelf, or smartphones that fell between couch cushions. The OCR capability reads prescription medication labels, cooking instructions on food packaging, and handwritten birthday cards. Additionally, the system can distinguish between similar canned goods in a pantry, identify clothing colors while shopping, detect obstacles like chairs in walkways, locate elevator buttons in unfamiliar buildings, identify currency denominations during transactions, and even recognize who has entered a room during social gatherings.

These capabilities directly address Seiple et al.'s (2025) findings by providing practical solutions that enhance independence in completing everyday tasks that typically require visual information. By leveraging these capabilities, we enhance independence and ease of access to important information.

### 3.3 Community-Driven and Familiar Integration

By fostering an open, extensible platform, we empower blind and low-vision users, developers, and accessibility experts to collaborate on tailored AI solutions. Research by Adnin and Das (2024) highlights that blind users create workarounds when using AI, underscoring the importance of community-driven innovation. Additionally, integrating with familiar and accessible platforms like WhatsApp reduces adoption barriers.

## 4. Limitations and Future Directions: Towards a More Robust Platform

While our project marks a significant step toward an extensible GAI platform for accessibility, several limitations remain. In its current form, the client is Windows-only and has no speech to text integration for interfacing. Future work will address these challenges to enhance the platform's usability and effectiveness.

### 4.1 Bring the Community Into the Picture

The capabilities of the current system are limited. But the extensibility promised by this system has great potential. To uncover that, we need to assess the opportunities and challenges that BVI hackers face when extending the system to meet their own visual accessibility needs. To gain this insight, we are developing a plan for a hackathon-style event with BVI developers focused on visual accessibility technologies.

### 4.2 Computational and Model Limitations

The performance of AI-driven features depends on available computational resources, which can impact responsiveness and accuracy. Furthermore, biases in AI models may affect reliability and fairness, as noted by Adnin and Das (2024), who found that blind users frequently encountered inaccuracies and needed verification methods. Additionally, Huh et al. 2023, and Glazko et al. 2023 signal the inability of BVI users to verify the correctness of outputs generated by VLMs, which are often known to hallucinate.

Future improvements include investigating edge AI deployment to reduce latency and enhance privacy, as well as fine-tuning vision-language models for specific accessibility tasks.

## 5. Conclusion

WhatsAI represents a significant step towards realizing the potential of generative AI to enhance accessibility for blind and low vision users. By providing an extensible platform that integrates with familiar communication tools and leverages state-of-the-art AI models, we aim to empower users and foster a community-driven approach to innovation in this critical domain. While acknowledging current limitations, our ongoing work focuses on creating a more robust, accessible, and user-centric platform that can transform how blind and low vision individuals interact with the visual world.